\documentclass[10pt]{article}
\usepackage{conf_iap}
\usepackage{graphics}
\def\frac{$''$\hspace*{-.1cm}}

\def\h2{H\,{\sc ii}}
\begin{document}
\heading{%
%
The youngest massive star clusters in the Magellanic Clouds
%
} 
\par\medskip\noindent
\author{%
M.~Heydari-Malayeri$^{1}$, V.~Charmandaris$^{2}$, L.~Deharveng$^{3}$,
M.R.~Rosa$^{4}$, \\D.~Schaerer$^{5}$, \& H.~Zinnecker$^{6}$
}
\address{Obs. de Paris, DEMIRM, 61 Ave de l'Observatoire, F--75014 Paris, France.}
\address{Astronomy Department, Cornell Univ., Ithaca, NY 14853, USA}
\address{Obs. de Marseille, 2 Place Le Verrier, F-13248 Marseille Cedex 4, France}
\address{STECF - European Southern Observatory,  D-85748 Garching, Germany}
\address{Obs. Midi-Pyrenees, 14, Avenue E. Belin, F-31400 Toulouse, France}
\address{Astrophysikalisches Institut Potsdam, D-14482 Potsdam, Germany}

\begin{abstract}
High resolution observations with HST have recently allowed us to
resolve and study several very tight clusters of newly born massive
stars in the Magellanic Clouds. Situated in an extremely rare category
of \h2 regions, being only 5$''$ to 10$''$ across and of high
excitation and extinction, these stars are just hatching from their
natal molecular clouds. Since the SMC is the most metal-poor galaxy
observable with very high angular resolution, this work may provide
valuable templates for addressing issues of star formation in the very
distant metal-poor galaxies of the early Universe.

\end{abstract}
\section{Introduction}

Our search for the youngest massive stars in the Magellanic Clouds
(MCs) started almost two decades ago on the basis of ground-based
observations \cite{first}. This led to the discovery of a distinct and
very rare class of \h2 regions in these galaxies, that we called
high-excitation compact \h2 ``blobs'' (HEBs). Contrary to the typical
\h2 regions of these galaxies which are extended structures with sizes
greater than 50\,pc, HEBs are an order of magnitude smaller having
diameters of less than about 3\,pc. HEBs are probably the final stages
in the evolution of the ultra-compact \h2 regions, whose Galactic
counterparts are detected only at infrared and radio wavelengths.

\section{Current HST Imaging and Spectroscopy}

The discovery of the very young and compact \h2 regions has entered a
new era with the high resolution capabilities of the HST. Our recent
HST observations \cite{N81,N88,N159,N11,N83}, reveal the stellar
content of these objects, so far out of reach from ground-based
telescopes, and indicate a turbulent environment typical of newborn
massive star forming regions.  Our findings suggest that the true
number of massive stars in the Magellanic clouds is underestimated
since a large number of them are hidden in unresolved small clusters,
and even though all compact \h2 regions belong to the general category
of HEBs they display several distinct characteristics (see Fig. 1).

\pagebreak
\begin{figure}[!h]
\centerline{\resizebox{11cm}{!}{\includegraphics{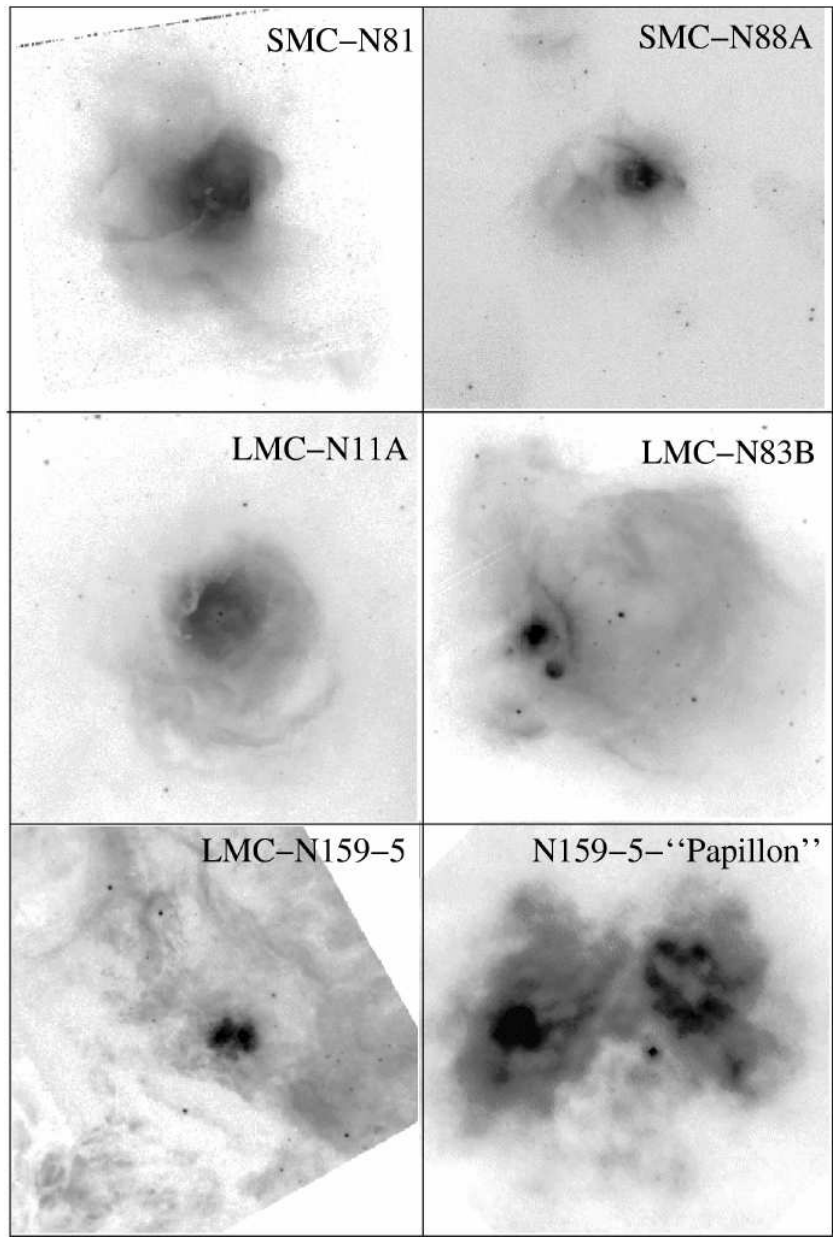}}}
\caption[]{
H$\alpha$ images of five of the HEBs observed by HST. North is up,
East is to the left and the field of view is 10$\times$10\,pc for each
image. A close-up view of the ``Papillon'', seen in N159-5 as a dark
spot above the absorption lane appearing in white, is presented as the
last image. See \cite{N81,N88,N159,N11,N83} for details.}
\end{figure}

A thorough study of the regions is presented in
\cite{N81,N88,N159,N11,N83}, but it is worth mentioning here a few of the
majors points. SMC N81, while young, is relatively more evolved than
N88A and also represents a really isolated massive starburst.  SMC
N88A is the latest starburst in a region where former generations of
massive stars are present over a large area. It has an elevated
extinction with values as high as A$_V$$\sim$3.5\,mag which is fairly
unique given the gas content and metallicity of the SMC. The LMC
region N159-5, similarly to SMC N88A, shows no conspicuous stars
probably due to its extreme youth.  Finally, LMC N83B represents a
rare opportunity where by means of combining the high resolution of
the HST with larger scale ground images a rather interesting spatial
distribution of the massive stars is observed. We find that this
distribution is consistent with the model of fractal / hierarchical
structure for the gas which gives rise to the star formation (see
\cite{N83}).

\begin{figure}[!ht]
\centerline{\resizebox{0.8\hsize}{!}{\includegraphics{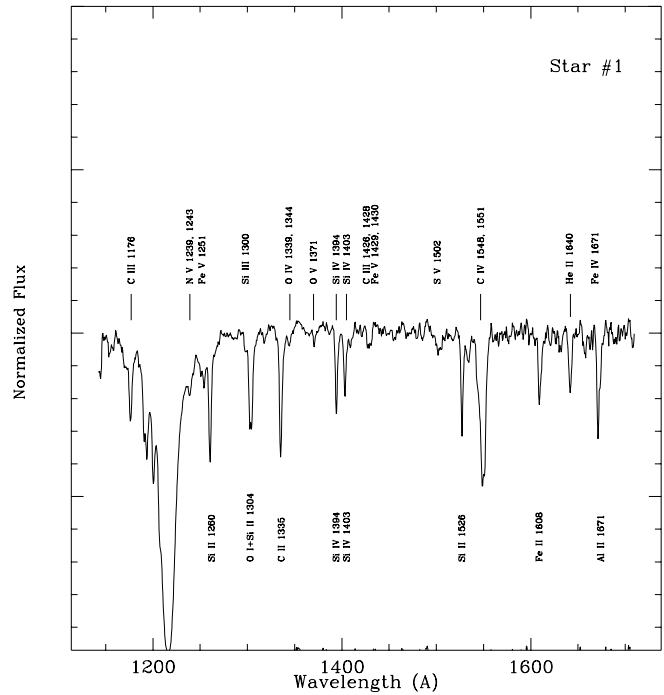}}}
\caption[]{
HST/STIS UV spectrum of the main exciting star in N81 (see \cite{STIS}). }
\end{figure}

To examine in more detail the properties of the young massive stars in
the HEBs we performed STIS spectroscopy of the major exctiting stars
of SMC N81 \cite{STIS}. The Far UV spectra (see Fig. 2), confirmed the
extreme youth we had inferred from their broad-band colors revealing
features characteristic of an O6--O8 stellar type. However, their
astonishingly weak wind profiles and their sub-luminosity (up to
$\sim$\,2 mag fainter in $M_{V}$ than the corresponding dwarfs) make
these stars a unique stellar population in the MCs. The weak wind must
be related to the low metallicity, since it is believed that the metal
deficiency of the SMC leads to a reduced radiation pressure
responsible for driving the winds of early type stars
(i.e. \cite{Leitherer92}). Our analysis suggests that these very
interesting stars are probably in the Hertzsprung-Russell diagram
locus of a particularly young class of massive stars, the so-called Vz
luminosity class (i.e. \cite{Walborn00}), as they are arriving on the
zero age main sequence.

\section{Perspectives}

Despite of the recent progress, several questions on the nature of
HEBs remain open. Is there an evolutionary secquence among HEBs? We
were able to identify some exciting stars in those regions, but we
also found evidence of high dust content and their absorption seems to
be ``patchy''. Could it be that our optical imaging misses a part of
an embedded the stellar population? Our data so far cannot exclude
the possibility of a low mass component in the luminocity function of
HEBs. However, if low mass stars were indeed absent, as the so-called
bi-model star formation theories predict \cite{Hans93}, this would
give further weight to coalescence scenarios for the formation of
massive stars (i.e \cite{Bonnell98}) according to which massive stars
form through collisions and coalescence of the low and intermediate
mass stellar component. Are the stellar winds of most stars in HEBs as
weak as the ones in N81, and if so, is this consistent with the fact
that the ionized gas surrounding them appears extremely turbulent with
strong ionization fronts, cavities, and bright gaseous filaments?

Clearly, high resolution near-IR imaging and spectroscopy of more
stars (currently attainable only with HST/STIS), would be highly
desirable in order to obtain sufficient new information to address
those issues.

\vspace*{-0.2cm}  

\vfill
\end{document}